\documentclass{appolb}
\usepackage{epsfig}
\begin{document}
\title{MARTINI - Monte Carlo simulation of jet evolution %
\thanks{Presented at the EMMI workshop and 26th Max-Born symposium in Wroclaw, Poland, July 9-11, 2009}%
}

\author{Bj\"orn Schenke, Charles Gale and Sangyong Jeon
\address{Department of Physics, McGill University, Montreal, Quebec, H3A\,2T8, Canada}
}
\maketitle
\begin{abstract}
We present the Modular Algorithm for Relativistic Treatment of heavy IoN Interactions (MARTINI),
an event generator for the hard and penetrating probes in high energy nucleus-nucleus collisions.
The simulation consists of a time evolution model for the soft background, such as hydrodynamics,
PYTHIA~8.1 to generate and hadronize the hard partons after the medium evolution, which is 
based on the McGill-AMY formalism and includes both radiative and elastic processes.
MARTINI allows for the generation of full event configurations in the high $p_T$ region.
We present results for the neutral pion and photon nuclear modification factor in Au+Au collisions at RHIC.
\end{abstract}
  
\section{Introduction}
High transverse momentum jets in heavy-ion collisions provide information on the produced hot quark-gluon plasma (QGP).
To extract this information from experimental data, it is important to develop a good theoretical understanding
of the particle production, the interactions of hard partons with the medium, the medium evolution, and the process of hadronization.

In the Modular Algorithm for Relativistic Treatment of heavy IoN Interactions (MARTINI) we incorporate 
all these aspects into a Monte Carlo framework, to create a most efficient connection between theory and experiment.
The creation of particles is taken care of by a slightly modified version of PYTHIA~8.1 that takes into account isospin effects.
Initial vacuum showering is also done by PYTHIA. MARTINI handles the subsequent medium evolution by sampling transition rates computed
using thermal field theory. Both radiative \cite{Arnold:2001ms, Arnold:2001ba, Arnold:2002ja} and elastic \cite{Schenke:2009ik}
processes are included. The transition rates depend on the thermal background, and information on the temperature, flow and QGP fraction  
is read in from external sources, like hydrodynamic simulation data. 
Finally, the evolved partons hadronize using PYTHIA's Lund model routines.

In this work we present results on neutral pion and photon production and compare to experimental data.

\section{The simulation}

At the core of MARTINI lies the McGill-AMY formalism for jet evolution in a dynamical thermal medium.
This evolution is governed by a set of coupled Fokker-Planck type rate equations of the 
form
\begin{eqnarray}\label{jet-evolution-eq}
\frac{dP(p)}{dt}\!=\!\int_{-\infty}^{\infty}\!\!\!\!\!\!dk
\left(\!P(p{+}k) \frac{d\Gamma(p{+}k,k)}{dk} - P(p)\frac{d\Gamma(p,k)}{dk}\!\right)\,,
\end{eqnarray}
where ${d\Gamma(p,k)}/{dk}$ is the transition rate for
processes where partons of energy $p$ lose energy $k$. 

In the AMY finite temperature field theory approach,
radiative transition rates can be calculated by means of integral equations
\cite{Arnold:2002ja}, which correctly reproduce both the Bethe-Heitler and the LPM results
in their respective limits \cite{Jeon:2003gi}. 
We include transition rates for the processes $g\rightarrow gg$, $q(\bar{q})\rightarrow q(\bar{q}) g$,
and $g\rightarrow q\bar{q}$. Furthermore, we include elastic processes employing the transition rates computed in
\cite{Schenke:2009ik}, and gluon-quark and quark-gluon conversion
due to Compton and annihilation processes, as well as the QED processes of
photon radiation $q\rightarrow q\gamma$ and jet-photon conversion.

\begin{figure}[tb]
  \begin{center}
    \includegraphics[width=8.5cm]{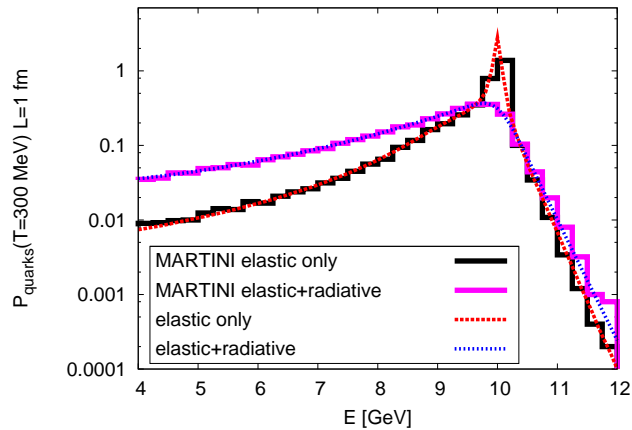}
    \caption{Quark distribution after passing through a brick of length $1\,{\rm fm}$ with temperature $300\,{\rm MeV}$.
      Comparing direct solutions of Eq.\,(\ref{jet-evolution-eq}) (dashed) to the MARTINI results. 
      The initial quark energy was $10\,{\rm GeV}$.
      See main text for details.}
    \label{fig:quarksElastic}
  \end{center}
\end{figure}

In MARTINI, we solve Eq.\,(\ref{jet-evolution-eq}) using Monte Carlo methods, keeping track 
of each individual parton, rather than the probability distributions $P$.
This way we obtain information on the full microscopic event configuration in the high momentum regime, including correlations,
which allows for a very detailed analysis and offers a direct interface between theory and experiment.
The average over a large number of events will correspond to the solution found by solving Eq.\,(\ref{jet-evolution-eq}) 
for the probability distribution. 
Fig. \ref{fig:quarksElastic} demonstrates this for a $T=300\,{\rm MeV}$ QGP brick of length $L=1\,{\rm fm}$. 
Here, we show the final quark distribution
after a quark of initial energy $E_i=10\,{\rm GeV}$ passed through the brick, comparing a direct solution of Eq.\,(\ref{jet-evolution-eq})
(see e.g. \cite{Schenke:2009ik}) and the result obtained after $10^5$ MARTINI runs, for only elastic and both radiative and elastic processes.

In a full heavy-ion event, the number of individual nucleon-nucleon collisions that produce partons with a certain minimal
transverse momentum $p_T^{\rm min}$ is determined from the total inelastic cross-section, provided by PYTHIA. 
The initial transverse positions of these collisions
are determined by the initial jet density distribution $\mathcal{P}_{AB}(b,\mathbf{r}_\perp)$, which is determined by the nuclear thickness
and overlap functions.
The initial parton distribution functions can be selected with the help of the Les Houches Accord PDF Interface (LHAPDF) 
\cite{Whalley:2005nh}.
We assume isospin symmetry and include nuclear effects on the parton distribution functions 
using the EKS98 \cite{Eskola:1998df} or EPS08 \cite{Eskola:2008ca} parametrization, by user choice.

The soft medium is described by hydrodynamics or other models, which provide information on the system's local temperature 
and flow velocity. 
Before the hydrodynamic evolution begins ($\tau<\tau_0$),
the partons shower as in vacuum. At this point, the AMY formalism does not include interference between vacuum and medium radiation.
So, we have explored two different implementations of the transition from vacuum to medium evolution. 
One is to include the complete vacuum shower, which is motivated by there being no apparent reason why the vacuum splittings should
end immediately once the medium has formed. Since most of the vacuum shower occurs before the medium has formed, this is a reasonable approximation. 
The other is to stop the vacuum evolution at the virtuality scale $Q_{\rm min}=\sqrt{p_T/\tau_0}$, determined by the
time $\tau_0$ at which the medium evolution begins. Using the latter method requires an about $10\%$ larger $\alpha_{\rm s}$ to describe the pion
$R_{AA}$ (see below).

During the medium evolution, individual partons move through the background according to their velocity.
Probabilities to undergo an interaction are determined in the local fluid cell rest frame using the transition rates and 
the local temperature.
If a process occurs, we sample the radiated or transferred energy from the transition rate of that process.
In case of an elastic process, we also sample the transferred transverse momentum, while
for radiative processes we assume collinear emission.

Radiated partons are also further evolved if their momentum is above a certain threshold $p_{\rm min}\simeq 2-3\,{\rm GeV}$. 
The overall evolution of a parton stops once its energy in a fluid cell's rest frame falls below the limit
of $4T$, where $T$ is the local temperature.
For partons that stay above that threshold, the evolution ends once they enter the hadronic phase of the background medium.
In the mixed phase, processes occur only for the QGP fraction.
When all partons have left the QGP phase, hadronization is performed by PYTHIA, to which the complete
information on all final partons is passed. Because PYTHIA uses the Lund string fragmentation model 
\cite{Andersson:1983ia,Sjostrand:1984ic}, it is essential to
keep track of all color strings during the in-medium evolution. 
For more information on the simulation please refer to Reference \cite{Schenke:2009gb}.

The concept of MARTINI is modular, such that we can turn on and off different processes independently, and use different hydrodynamic
or other data inputs.

\section{Results for one-body observables}
\begin{figure}[t]
  \begin{minipage}[t]{0.5\linewidth}
    \centering
    \includegraphics[width=6.5cm]{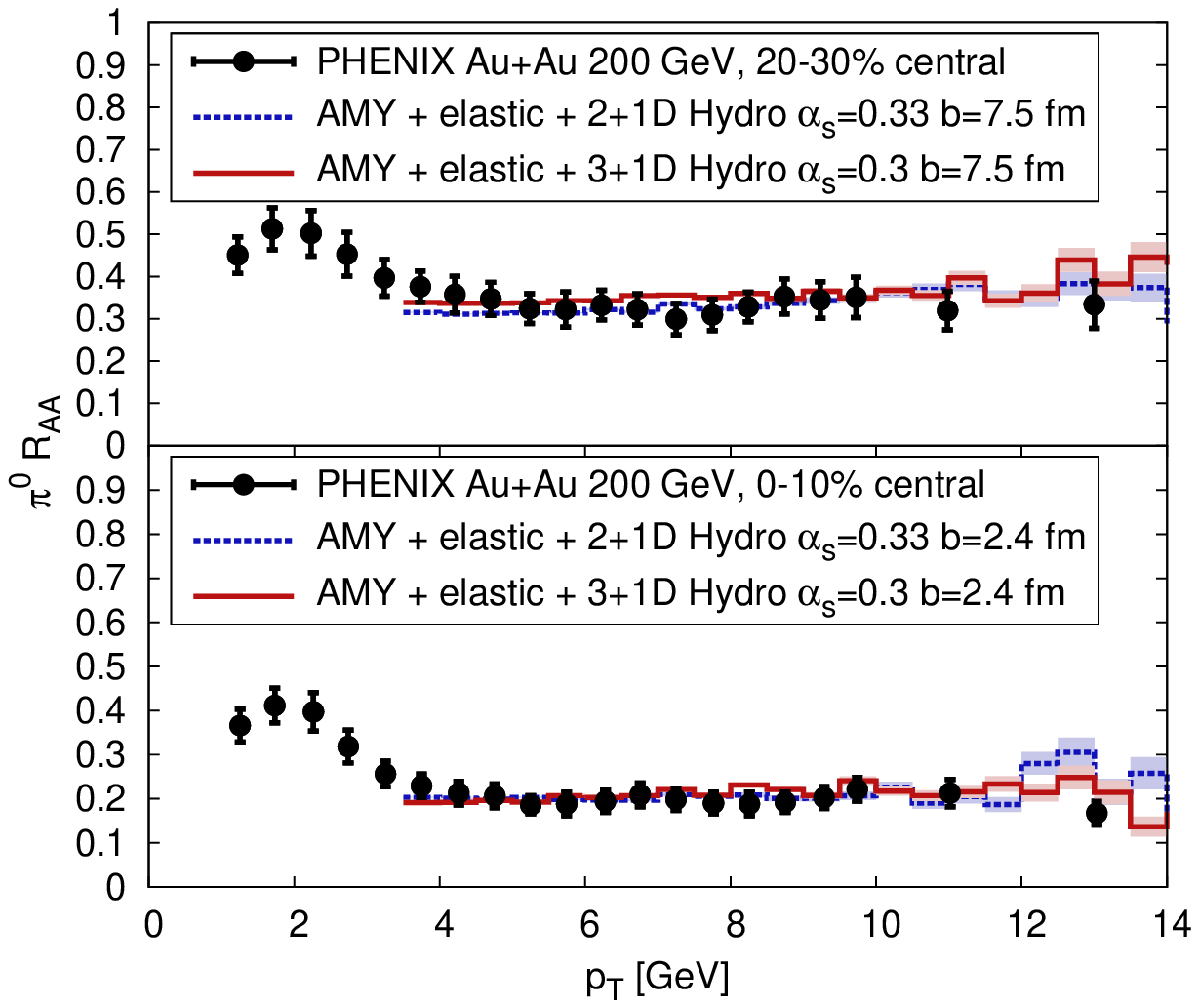}
    \caption{The neutral pion nuclear modification factor for mid-central (upper panel) 
      and central (lower panel) collisions at RHIC with $\sqrt{s}=200\,{\rm GeV}$.
      MARTINI results ($b=2.4\,{\rm fm}$ and $b=7.5\,{\rm fm}$) using 3+1 \cite{Nonaka:2006yn} (solid) and 2+1 \cite{Kolb:2000sd,Kolb:2002ve,Kolb:2003dz} (dashed) dimensional hydro evolution compared to PHENIX data from \cite{Adare:2008qa}.}
    \label{fig:raa}
  \end{minipage}
  \hspace{0.5cm}
  \begin{minipage}[t]{0.5\linewidth}
    \includegraphics[width=6.5cm]{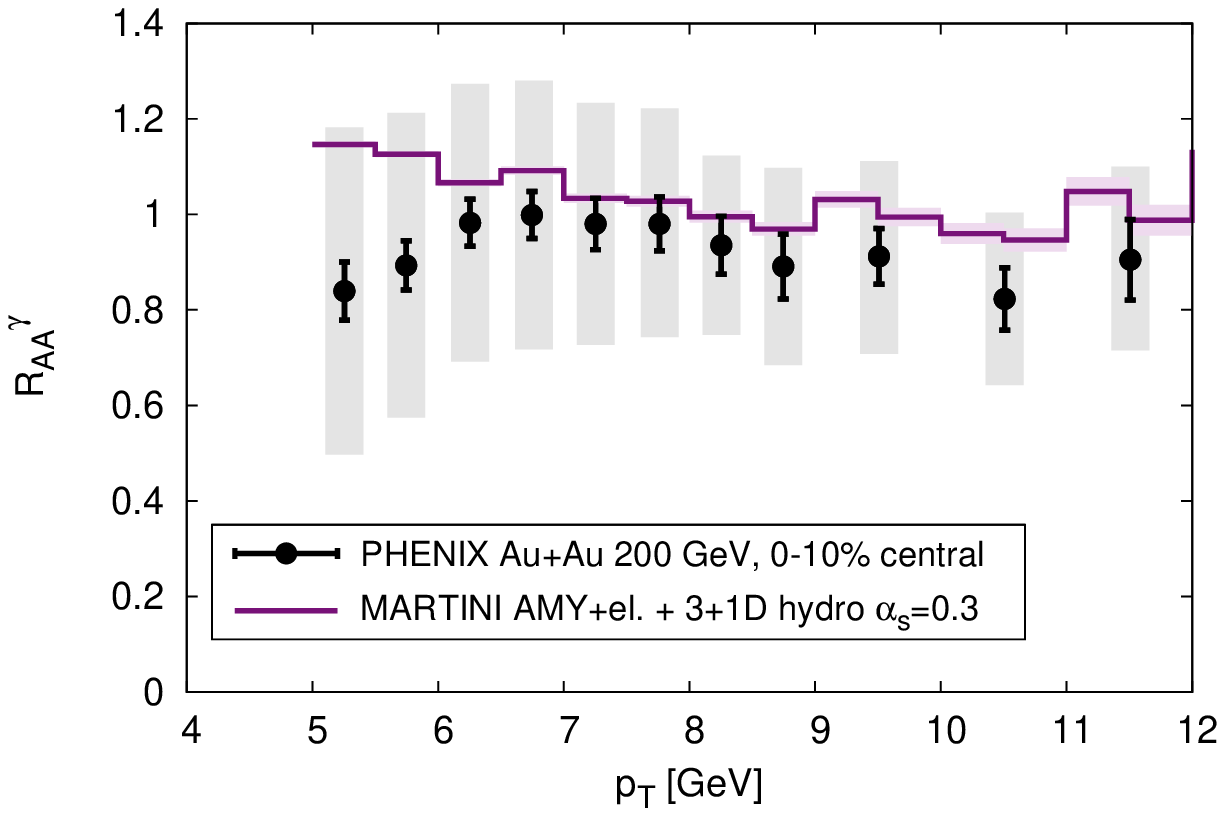}
    \caption{(Color online) Photon nuclear modification factor for central collisions at $\sqrt{s}=200\,{\rm GeV}$.
      MARTINI results compared to data from \cite{Isobe:2007ku}. See main text for details.}
    \label{fig:photon-raa}
  \end{minipage}
\end{figure}

In Fig.~\ref{fig:raa} we present the results for the neutral pion nuclear modification factor, defined by
\begin{equation}
  R_{AA}=\frac{1}{N_{\rm coll}(b)} \frac{dN_{AA}(b)/d^2p_T dy}{dN_{pp}/d^2p_Tdy}\,,
\end{equation}
in Au+Au collisions at RHIC measured at mid-rapidity in two different centrality classes (0-10\%) and (20-30\%),
employing the corresponding average impact parameters, $2.4\,{\rm fm}$ and $7.5\,{\rm  fm}$.
Au+Au calculations take into account both radiative and elastic processes in the medium described by 
either the 2+1 dimensional hydrodynamics of \cite{Kolb:2000sd,Kolb:2002ve,Kolb:2003dz} 
or the 3+1 dimensional hydrodynamics of \cite{Nonaka:2006yn}, using a coupling constant $\alpha_{\rm s}=0.33$ or
$\alpha_{\rm s}=0.3$, respectively. 
$\alpha_{\rm s}$ was adjusted to describe the experimental measurement of the neutral pion nuclear modification factor
$R_{AA}$ in most central collisions.
The same value of $\alpha_{\rm s}$ is used in all following calculations. 
We find very good agreement with the data for both centrality classes.

We also studied photon production within MARTINI.
Apart from direct photons and those produced in the PYTHIA showers, 
MARTINI includes jet-medium photons from photon radiation 
and jet-photon conversion.
Fig.~\ref{fig:photon-raa} shows $R_{AA}^\gamma$ as a function of $p_T$ for most central Au+Au collisions 
($b=2.4\,{\rm fm}$) at RHIC compared with $0-10\%$ central PHENIX data. 
The presented result includes all the vacuum final state radiation and is hence an upper limit for $R_{AA}^\gamma$.
Including interference between medium and vacuum radiation is a future task.

For more detailed results please refer to \cite{Schenke:2009gb}.

\section{Conclusions and Outlook}
\label{conclusions}
We presented first results obtained with the newly developed
Modular Algorithm for Relativistic Treatment of heavy IoN Interactions (MARTINI).
This hybrid approach describes the soft background medium using hydrodynamics or other medium models
and simulates the hard event microscopically, using PYTHIA~8.1 to generate individual hard
nucleon-nucleon collisions. Hard partons are evolved through the medium using the McGill-AMY
evolution scheme including radiative and elastic processes.
On the parton level, we found the same result as a direct solution of the Focker-Planck type rate equations for the brick problem.
Fragmentation is performed employing PYTHIA~8.1 which uses the Lund string fragmentation model. 
Apart from parameters in PYTHIA which were fixed by matching the neutral pion and photon spectra in p+p collisons
to experimental data, $\alpha_{\rm s}$ is the only free parameter.
Employing the 3+1 dimensional hydrodynamic evolution from \cite{Nonaka:2006yn}, it was set to
$\alpha_s=0.3$ (and $0.33$ for the 2+1 dimensional hydro) to match the neutral pion $R_{AA}$ measurement for central collisions.
Using the same value for all other calculations (there was no additional freedom in any of the calculations), 
we were able to describe the neutral pion $R_{AA}$ in mid-central collisions as well as the photon $R_{AA}$ in the
regarded $p_T$ range. 

We showed that MARTINI can reproduce one-body observables in good agreement 
with the data. The next step will be to explore its full potential 
by studying many-body observables and correlations.
Another future task is the implementation of heavy quark evolution.

\section*{Acknowledgments}
\hyphenation{Abhijit Majumder Ulrich Heinz Michael Strickland Richard Tomlinson Steffen Bass Evan Frodermann
Chiho Nonaka Huichao Song Guy Moore Guang Qin Vasile Topor-Pop}
We are happy to thank Steffen Bass, Evan Frodermann, Ulrich Heinz, Scott Moreland,
Chiho Nonaka, Torbj\"orn Sj\"ostrand, and Huichao Song 
for very helpful correspondence. We thank
Guy Moore, Guang-You Qin, Thorsten Renk, Michael Strickland, and Vasile Topor-Pop for helpful discussions and comments. 
This work was supported in part by the Natural Sciences and Engineering Research Council of Canada. 
B.S.\ gratefully acknowledges a Richard H.~Tomlinson Fellowship awarded by McGill University.
\bibliography{martini}

\end{document}